\documentclass{svproc}
\usepackage{natbib}			
\usepackage{mnemonic}			

\newcommand{\kms}{\,km\,s$^{-1}$}

\begin{document}

\mainmatter
\title{Gaia, Fundamental Physics, and Dark Matter}
\titlerunning{Gaia and Dark Matter}  
\author{Michael Perryman\inst{1} \and Konstantin Zioutas\inst{2}}
%
\authorrunning{Perryman \& Zioutas} 
\institute{
\inst{}University College Dublin
	\and
\inst{}University of Patras, Patras, Greece
}

\maketitle

\centerline{\small Invited contribution:}
\centerline{\small 16th Patras Workshop on Axions, WIMPs and WISPs, 14--18 June 2021\footnote{\tiny https://agenda.infn.it/event/20431/contributions/137758/attachments/82513/108430/gaia-patras-2021.Perryman.pdf}}

\begin{abstract}
The Gaia space astrometry mission is measuring accurate distances and space motions of more than two billion stars throughout our Galaxy and beyond. This is a first look at how Gaia is contributing to fundamental physics, and in particular to our understanding of dark matter, for which a few examples are given from the current literature. 

One of our goals is to illustrate how deep, and often surprising, insight into very diverse areas of fundamental physics can be extracted from this new and enormous high-accuracy stellar data set.  In this spirit, we finish by suggesting a search for a connection between stellar activity, dark matter streams, and planetary configuration in nearby exoplanetary systems, as has been tentatively proposed in the case of the solar system. 

\end{abstract}

\section{Introduction}
\subsection{Gaia in brief}
Gaia is a scientific satellite of the European Space Agency, launched in 2013, and seven years in to a possible 10-year operational lifetime. Building on the success of its predecessor Hipparcos, the high-accuracy telescope slowly and continuously scans the celestial sphere from its deep L2 Langrangian orbit, and sends to the ground a high-data rate stream of CCD focal-plane measurements which encode the along-scan positional differences between the stars observed at that measurement epoch. 

Due to the fact that all stars are moving through space, as well as appearing to move as a result of the Earth's orbital motion around the Sun, repeated measurements in different scanning orientations over several years provides a system of deeply interconnected observations from which the distances and space motions can be derived.

\subsection{A concise history of astrometry}
Positional astronomy (or astrometry) has a long history dating back over 2000 years to the first enquiries by the Ancient Greeks. Many deep insights about our Galaxy and the Universe have been uncovered along this journey, and a few highlights are mentioned here to set the scene.

The resurgence of scientific enquiry in Europe around 1500--1700 led to the erroneous Earth-centred view of the Universe being eventually replaced by Copernicus's heliocentric model in which the planets orbit the Sun in elliptical orbits, rather than the planets and the Sun orbiting the Earth in circular epicycles. Better observations, for example by Tycho Brahe, aided the formulation of Kepler's laws of planetary motion, and the advances of Newton's laws of gravity and motion. As positional measurements improved down the centuries, in part driven by the requirements for accurate navigation at sea, other discoveries followed.

In 1718, Edmond Halley deduced the stars were moving through space. 
In 1725, James Bradley measured the effect of stellar aberration, concluding that the Earth was moving through space, and that the speed of light was finite.
In 1783, William Herschel inferred that the Sun itself was moving through space.
 
In 1838, Friedrich Bessel, Otto Struve and Thomas Henderson, independently measured the first star distances through the effect of parallax (as the Earth orbits the Sun), in the process providing direct proof of the immensity of space.

\subsection{A note on angles and distances}
Star positions, and their inferred motions, are measured as differential angles on the celestial sphere. To give some context, the measurements of the ancient Greek astronomer Hipparchus were at the level of a few minutes of arc. Tycho Brahe, around 1600, measured 1000 stars with an accuracy of better than 100~arcsec, while Flamsteed, around 1700, measured some 4000 stars to around 10~arcsec. This can be compared to the resolution limit of the unaided human eye, around 30~arcsec. By the end of 20th century, the most accurate ground-based catalogues, of some 1400 stars (the FK5) reached a few tenths of an arcsecond. But the Earth's turbulent atmosphere imposes a limit on individual observations of around 1~arcsec or a little better, which ingenuity consistently failed to overcome.

\subsection{The move to space}
Because the nearest stars have a parallax of about 1~arcsec (being the distance at which the Earth's orbit subtends this angle), determining stellar distances required the measurements of much smaller angles. This, in turn, eventually led to the first space mission, ESA's Hipparcos, operated between 1989--1993, dedicated to the accurate positional measurements of stars. Its catalogue of 120\,000 stars with accuracies of 1~milli-arcsec was published in 1997. This positional accuracy corresponds to distances to 10\% accuracy at 100~parsec. 

Hipparcos was a huge advance, and established the principles of carrying out astrometric measurements from space. But a distance of 100~pc is small compared to the size of our Galaxy (the Sun lies at distance of about 8\,000~parsec, or around 30\,000 light-years from its centre). And the Hipparcos catalogue of 120\,000 stars is but a tiny sample of our Galaxy's estimated 100~billion stars.

The Gaia mission followed fairly quickly. Accepted by ESA in 2000, and launched in 2013, it is surveying more than two billion stars down to 21~magnitude, yielding accuracies of 20~micro-arcsec at 15~mag; somewhat better for brighter stars, and somewhat less for fainter stars. These distance and space motion measurements are revolutionising the knowledge of the distances, motions, and physical properties of a major fraction of our Galaxy's stellar population. Stars in our neighbouring Local Group galaxies are being measured in huge numbers as well.

For those looking in from outside of astronomy, it may seem surprising that Gaia's impact on so many aspects of astronomy is so vast. For the most part, this rests on the crucial use of stellar distances to transform observed quantities, such as a star's apparent brightness or angular radius, into the more physically meaningful quantities such as absolute luminosity and radius. These provide powerful constraints on highly sophisticated stellar evolution models, and in turn on properties such as mass and age.

\section{Applications to fundamental physics}
Various applications of the Gaia data to questions in fundamental physics were considered during the preparatory phase of the mission. Although new results on some of these are not yet available, we expect them to be quantified in due course. Just a few examples are given below.

\subsection{Gravitational light bending}
The reduction of the Hipparcos data required the inclusion of stellar aberration up to terms in $(v/c)^2$, and the general relativistic treatment of light bending due to the gravitational field of the Sun (and Earth). 

The Gaia reduction requires a more accurate and comprehensive inclusion of relativistic effects, at the same time providing the opportunity to test a number of parameters of General Relativity in new observational domains, and with much improved precision. No results are yet available, but the possibility of accurately measuring the parameter $\gamma$ of the Parameterised Post-Newtonian (PPN) formalism is of key importance in fundamental physics, and it will be eventually determined by Gaia with a very high accuracy.

Beyond the solar system, gravitational light bending is an important effect in interpreting the structure of quasars observed with Gaia \citep[e.g.][]{2017MNRAS.472.5023L,2018MNRAS.479.5060L, 2019MNRAS.483.4242L, 2018A&A...618A..56D}.
Gravitational lensing has found applications in constraining the mass of MACHOs in the Galactic halo, and is today enjoying success in the discovery and characterisation of planetary systems. In addition to the photometric effects, astrometric effects are expected to be observable in a number of cases
\citep[e.g.][]{2018A&A...618A..44B,2018A&A...615L..11K, 2018A&A...617A.135M}.

\subsection{Perihelion precession and solar quadrupole}
Gaia is observing and discovering many tens of thousands of minor planets \citep[e.g.][]{2018A&A...616A..13G,2018AJ....156..139M}.  Most of these will belong to the asteroidal main belt, with small orbital eccentricity and semi-major axes close to 3~au, a circumstance not favourable to detect significant relativistic effects in their orbital motion. Effects are stronger for the Apollo and Aten groups, which are all Earth-orbit crossers, and include objects with semi-major axes of the order of 1~au and eccentricities as large as~0.9.  

The relativistic effect and the solar quadrupole cause the orbital perihelion of a solar system body to precess at a rate which can be characterised in terms of the PPN parameters. For the main belt, the precession is about seven times smaller than for Mercury  in rate per revolution, although more than a hundred times in absolute rate. It is has been estimated that the PPN precession constant, $\lambda = (2\gamma -\beta +2)/3$, should be determined with an accuracy of at least $10^{-4}$.

\subsection{Secular change of the gravitational constant}
The possibility of a time variation of the constant of gravitation, $G$, was first considered by \citet{1938RSPSA.165..199D} on the basis of his large number hypothesis, and later developed by Brans \& Dicke (1961) in their theory of gravitation. Variation could be related to the expansion of the Universe, in which case $\dot G/G=\sigma H_0$, where $H_0$ is the Hubble constant, and $\sigma$ is a dimensionless parameter whose value depends on both the gravitational constant and the cosmological model considered (Will 1987), with the standard model having $G=$~constant and $\sigma=0$. Revival of interest in the Brans-Dicke-like theories, with a variable $G$, was partially motivated by the appearance of superstring theories where $G$ is considered to be a dynamical quantity, and in which a scale-dependent gravitational constant could mimic the presence of dark matter \citep{1992PhLB..281..219G}.

Of various routes by which Gaia might contribute, white dwarfs provide a well-studied possibility \citep[e.g.][]{1995MNRAS.277..801G}. This is in part because, when they are cool enough, their energy is entirely of gravitational and thermal origin, and any change in $G$ modifies the energy balance which in turn modifies the luminosity. Furthermore, since they are long-lived objects, with life-times of order 10~Gyr, even extremely small values of $\dot G$ can become prominent.

\subsection{Hypervelocity stars as evidence for supermassive black holes}
Stars in our solar neighbourhood have space velocities of typically 20--30\kms\ with respect to our Galaxy as a whole. Their velocities arise from the conditions of their birth in star clusters and star associations, and these are later modified by interactions with other mass structures of the Galaxy. 

`Runaway stars', with much higher velocities, of 100\kms\ or more, have been known since the 1960s. They acquired these velocities from binary supernova ejections, or dynamical ejections from star clusters, in which extreme gravitational encounters hurl one of the interacting stars outwards at high speed. 

Hypervelocity stars are an extreme type of runaway star which originate near a supermassive black hole. In this predicted Hills ejection mechanism \citep{1988Natur.331..687H}, the black hole acts as a gravitational slingshot.  Simulations show that stars can be ejected from the deep potential well of a massive black hole, either as a result of scattering with another star, or through tidal breakup of a binary star system.  In the process, stars could be ejected at 1000--2000\kms\ or more, greatly exceeding speeds that could arise from either binary supernova or star cluster mechanisms. The most extreme velocity objects would not be gravitationally bound to our Galaxy, and will escape the confines of even its vast boundaries after some 300~Myr.

Since the DR2 results were made available in April 2018, Gaia is providing proper motions and distances of previously known hypervelocity star candidates with unprecedented accuracies, in turn helping to pinpoint their origin, while also searching the entire sky for other examples of these very rare objects. 

Amongst new discoveries from the Gaia DR2 data release, a number are believed to originate from the supermassive black hole at the Galactic centre, while others do not \citep[e.g.][]{2018AJ....156...87L, 2019ApJS..244....4D}.

As well as probing conditions close to the central black hold, the Gaia results on hypervelocity stars could provide a significant ingredient for refining cosmological models. In the `concordance $\Lambda$CDM model', galaxies are embedded within extended halo structures, largely made of some non-viscous dark matter, visible only through its gravitational effects. Over cosmic time, haloes grow in mass and size through hierarchical clustering, starting from the initial perturbations of a slightly inhomogeneous matter density field, but with resulting shapes and masses depending on the model details. 

As a result of the enormous distances travelled on their journey through the Galaxy halo, in different directions and now at a vast range of distances from their origin, they probe the Galaxy's gravitational potential, making them possible tracers of the detailed (dark) matter distribution in the Milky Way \citep[e.g.][]{2005ApJ...634..344G,2006ApJ...651..392S}.

\subsection{Ultra-wide binaries as tests of gravity}
Binary, and occasionally triple or even higher multiplicity star systems, form -- over a range of separations -- in the swirling gas clouds of dense regions of the interstellar medium, such as density enhancements triggered by the passage of rotating spiral density waves. If a binary forms with a small separation, their orbits can slowly spiral inwards until they eventually coalesce. Wider separation binaries can be slowly torn apart by external gravitational forces. 

An ultra-wide binary with a separation of 0.5~parsec (1.6~light-years, or 100\,000 astronomical units) is likely to break up within about 100 million years \citep{1937AZh....14..207A}. A binary with a separation of 0.1~pc (0.3~light-years) might survive for more than a billion years.
At these enormous separations, two stars will be widely separated on the sky, with very long orbital periods, but sharing an almost identical space motion over millennia. Until Gaia, physical binaries with ultra-wide separations have been extremely difficult to identify. 

With their identification, new tests of gravity become available. For example, the original MOND-like models result in relative orbital velocities which are  significantly different to those predicted by Newtonian models, specifically they can allow bound binaries with relative velocities well above the Newtonian `ceiling'. 

Presently, wide-separation binaries from Gaia have relative velocities well above those permitted by Newtonian theory, but it appears that the distribution tail can be explained by pairs of stars which were born in the same open cluster, but which are currently undergoing a chance close `flyby' \citep[e.g.][]{2019MNRAS.488.4740P,2019IJMPD..2850101H,2020MNRAS.491L..72C}.

\subsection{Measurement of gravitational redshift}
Gaia can measure the gravitational redshift induced by the extreme surface gravity of highly compact white dwarfs. Some background is needed.
 
Classical astrometry ignores a star's radial velocity, i.e.\ its space motion {\it along\/} the line-of-sight. This is because a star's radial velocity generally has no effect on its position on the sky. But radial velocities are no longer irrelevant in very high accuracy astrometry, where it can affect the determination both of the parallax of a star, and its proper motion. For the latter, the tangential velocity changes due to the varying angle between the line-of-sight and the direction of its space velocity.  The two effects result in a changing proper motion with time, which is interpreted as an apparent (or `perspective') acceleration of the star's motion on the sky, and which is proportional to the product of the star's parallax, its proper motion, and its radial velocity.  It is always a tiny effect, but largest for nearby stars with a high proper motion, and the phenomenon is being well characterised with Gaia \citep{2016A&A...595A...4L,2018A&A...616A...2L,2021A&A...649A...2L,2021arXiv210509014L}.

The reason that this is of interest is that classical (spectroscopic) Doppler velocity measurements yield the combination of the true velocity of the star's barycentre, combined with surface effects such as atmospheric dynamics and gravitational redshifts. The determination of stellar radial velocities from geometric principles, i.e.\ without using spectroscopy or invoking the Doppler principle, can then be used, in principle at least, to examine these stellar phenomena. 

Amongst these are the possibility of measuring the gravitational redshifts of white dwarfs. To date, only in the nearby system of Sirius~B has a precise value of $80.65\pm0.77$\kms\ been measured (by Hubble Space Telescope). With Gaia, the most precise determination of this astrometric radial velocity for white dwarfs so far is for LAWD~37 ,with $28\pm5$\kms\ \citep{2021arXiv210509014L}, although this object does not yet have a published spectroscopic radial velocity, so its gravitational redshift can not yet be inferred.

\section{Phenomena in the context of $\Lambda$CDM}

\subsection{Dark matter in the Galactic plane}
Stars rotate around the Galaxy with an orbital period of about 250~Myr. 
Their vertical velocity component is controlled by the mass density of the Galactic plane, and stars oscillate about the plane with a period of order 50~Myr.
The vertical force acting on the stars can be reconstructed from the vertical velocity dispersion, and the vertical density profile, both of which require good estimates of the distances and space velocities of a suitable tracer star population.  Once this is estimated, a comparison with the local mass density of visible matter yields any dark matter contribution.

Results from Hipparcos \citep[e.g.][]{1998A&A...329..920C,1999A&A...341...86B} yielded a picture in which disk matter is well accounted for (by stars, and interstellar gas and dust), and in which dark matter in the Galaxy must be distributed in the form of the spherical halo, with little if any concentrated in the form of the disk.  Much more will come from Gaia about the sub-structure of the Galaxy disk and the Galaxy halo.

\subsection{Galactocentric acceleration of the Sun}
The motion of stars around the Galaxy, results in an acceleration in the motion of the Sun. Despite the enormously long orbital period of 250~Myr, this acceleration has been detected in the velocity field of quasars by Gaia, and leads to a fundamental interpretation in terms of the motion of our own Galaxy with respect to the Local Group of galaxies, and with respect to the Cosmic Microwave Background \citep{2021A&A...649A...9G}.

\subsection{Dwarf spheroidal galaxies}
Dwarf spheroidal galaxies, or dSph, are small, low-luminosity galaxies comprising an old stellar population with very little dust. In contrast to dwarf elliptical galaxies, they are roughly spheroidal in shape. Some two dozen are known as companions to either the Milky Way or to the Andromeda Galaxy (M31). They are named after the constellation in which they are found.
The first known, Sculptor and Fornax, were discovered by Harlow Shapley in 1938, where he described them as {\it `unlike any known stellar organisation'}. But despite weighing in at around $10^7$ solar masses, further discoveries were challenged by their low luminosities and low surface brightnesses. 

By the late 1990s, their rarity seemed to be in conflict with the $\Lambda$CDM cosmological model, which predicted that massive galaxies like the Milky Way should be surrounded by many dark matter dominated satellite halos.  This conflict eased with the discovery of around a dozen very faint Local Group dwarfs from the Sloan Digital Sky Survey around 2000, and a similar number discovered by the Dark Energy Survey around 2015. 

These dwarf spheroidal galaxies are typically very distant, ranging from about 26~kpc in the case of Sagittarius, to around 250~kpc in the case of Leo~I, well beyond the Magellanic Clouds. Their bulk proper motions are consequently very small, rarely reaching 0.5~milli-arcsec per year. The best previous determinations of these tiny motions have mostly been made possible from the Hubble Space Telescope. 

The motions of 12 of these, from Gaia, have been studied by \citet{2018A&A...616A..12G}, who then used their derived positions and space motions, along with various state-of-the-art models of the Galaxy's mass distribution (for example comprising a stellar bulge, star and gas disks, and a dark matter halo), to follow their orbits backwards in time around the Galaxy over the past 250~million years. 

Most, it turns out, are on (slightly) prograde orbits, while Fornax is retrograde, qualitatively similar to what has been found for globular clusters. However, their orbital eccentricities are very different. Few have very eccentric orbits, with Carina even somewhat circular.

All this leads to two important conclusions. First, there is only a weak similarity, if any, between the orbits of globular clusters and dwarf spheroidals. Second, their eccentricity distribution is inconsistent with the findings of recent cosmological simulations, where they are predicted to be on rather radial orbits. This ordered complexity might indicate some collective infall from a preferred direction, perhaps a `cosmic web filament' aligned with the Galactic $z$-axis. But it appears to exclude a single event underlying their origin.

\subsection{Galaxy mergers}
Models of structure formation in the Universe suggest that our Galaxy's inner stellar halo should be dominated by the debris of just a few massive progenitor galaxies merging with our own early on in its formation history. This explanation finds convincing support with the new Gaia data, where there is compelling fossil evidence for one such merger event which took place around 10~Gyr ago, and whose debris in the solar neighbourhood reveals much about this enormous and disruptive event.

A study by \citet{2018Natur.563...85H} demonstrated that the inner halo is dominated by debris from an object which at infall was slightly more massive than the Small Magellanic Cloud, and which the authors referred to as Gaia--Enceladus.

The stars originating from the Gaia--Enceladus accretion event cover nearly the entire sky, and their motions reveal the presence of streams and slightly retrograde and elongated trajectories. Hundreds of RR~Lyrae stars and thirteen globular clusters following a consistent age--metallicity relation can be associated to the merger on the basis of their orbits.  With an estimated 4:1 mass-ratio between the young Milky Way and Gaia--Enceladus, the merger would have led to the `dynamical heating' of the precursor of the Galactic thick disk, increasing their space velocities, and therefore increasing their scale height with respect to the Galaxy mid-plane. It seems highly plausible that the merger between Gaia--Enceladus and the Milky Way contributed to the formation of our Galaxy's thick disk component some 10~Gyr ago.  And most probably, this was the last significant merger that our Galaxy experienced.

Amongst these very large-scale cosmological simulations, the EAGLE project has been shown to produce a realistic population of galaxies reproducing a broad range of observed galaxy properties. The largest of the EAGLE simulations, L100N15043, has a cubic volume of 100~Mpc in size, and includes the effects of both baryonic and dark matter. Its huge volume of simulated space--time provides numerous Milky Way-type galaxies, and with a wide range of merger histories.
Amongst these mergers, \citet{2019ApJ...883L...5B} identified one with remarkably similar properties to the Gaia--Enceladus event, also occurring around 9~Gyr ago. These specific  simulations result in merger debris on a slightly retrograde orbit (as found for Gaia--Enceladus), bursts of star formation in the early disk, the formation of a dynamically heated thick disk (as seen in our Milky Way), and with a large fraction of the debris deposited at large heights above the Galactic disk, corresponding to our Milky Way's stellar halo.
All-in-all, a most remarkable triumph of state-of-the-art space observations combined with state-of-the-art cosmological simulations!

\subsection{Tumbling of the Milky Way Galaxy}
The Gaia measurements provide an extremely rigid set of star positions, parallaxes, and space motions. But since all stars are measured with respect to other stars, there turns out to be an undetermined global rotation of the reference frame with respect to an inertial reference system, with three degrees of rotational freedom, along with a further three degrees of freedom in the `spin', or angular rotation rate, of the resulting reference frame. By design, the Gaia observations extend to such faint magnitudes, around 21~mag, that there are some 500\,000 distant quasars directly observable. These establish both the rotation and spin of the Gaia reference frame, providing the resulting stellar motions within an inertial reference system \citet{2018A&A...616A..14G}.

Such an accurate reference frame may have cosmological implications previously considered unimportant and unmeasurable, such as detecting the tumbling of our Galaxy's triaxial dark matter halo  
\citep{2014ApJ...789..166P}.

\section{The physics of stars}
This is a huge topic for Gaia, and we give just a few examples of some important physics that is being extracted from the data. 

Many insights make appeal to the Hertzsprung--Russell diagram, which shows the relationship between a star's luminosity (or absolute magnitude) versus its temperature (or colour). It was created independently around 1910 by the Ejnar Hertzsprung and Henry Norris Russell (physicists will know his name in the context of quantum mechanical `LS coupling' or Russell--Saunders coupling). This graphical presentation of stellar properties, in which stars of higher luminosity are towards the top of the diagram, and stars with higher surface temperature (or bluer colour) are towards its left side, facilitated a major step in understanding stellar evolution.

The Hertzsprung--Russell diagram remains a powerful tool for interpreting, through stellar evolutionary models, the properties of individual stars, star clusters, and entire stellar populations.
The fundamental relevance of Gaia here is that accurate parallaxes, especially when combined with the simultaneous accurate Gaia photometry, then yield accurate absolute magnitudes, which allow them to be precisely located in the diagram. 

\subsection{The internal structure of M~dwarfs}
While the main sequence in the HR diagram is generally smoothly populated as a function of stellar temperature, a tiny but pronounced discontinuity in number density occurs within the sequence of the coolest, faintest red dwarfs. This gap was first reported in the Gaia data by \citet{2018ApJ...861L..11J}.  Qualitatively, the gap arises because of the very different internal structure of stars on either side of it: low-mass M~dwarfs have a fully convective interior, while more massive stars (including the Sun) have a convective envelope surrounding a radiative zone. 
The transition occurs at about $0.35M_\odot$.

The gap arises because the convective motions in the cores of low-mass stars help mix the intermediate nuclear fusion products, allowing the star to fuse hydrogen more efficiently. 
Detailed models show that the mixing of $^3$He during the merger of the envelope and core convection zones occurs over a narrow range of masses, successfully replicating an associated dip in the luminosity function which is responsible for the gap \citep{2018MNRAS.480.1711M}.

\vspace{-5pt}
\subsection{Core crystallisation in white dwarfs}
In the lower-left part of the HR~diagram, the white dwarf sequence shows several remarkable features, identified by \citet{2018A&A...616A..10G}. They constructed a Gaia sample with relative parallax uncertainties better than 5\%, yielding a set of 26\,264 white dwarfs.
In addition to a prominent split in the white dwarf cooling sequence between H and He white dwarfs, there is a weaker concentration to the hotter side of the white dwarf sequence \citep{2019Natur.565..202T}.  This has been attributed to core crystallisation as the white dwarfs cool. 
In the process, the hot plasma fluid (of nuclei and electrons) releases an associated latent heat, providing a new source of energy that delays the object's cooling. 

Gaia's white dwarfs thus provide direct evidence that a first-order phase transition occurs in high-density Coulomb plasmas. It is a theory that cannot be tested in terrestrial laboratories due to the extreme densities involved.

\vspace{-5pt}
\section{Stars and planets as dark matter detectors}
We finish with a recent proposal regarding the use of stars and planets as dark matter detectors, through possible effects on stellar activity.
 
So far, there are two main hypotheses for a mechanism underlying the solar cycle, both driven by solar axial rotation: it may originate from a turbulent dynamo operating in or below the convection envelope, or from a large-scale oscillation superimposed on a fossil magnetic field in the radiative core. 

However, the precise dynamo nature, and many details of the associated solar activity (such as the details of the sun spot cycles, or the prolonged Maunder-type solar minima) remain unexplained. While planetary {\it tidal\/} forces have been variously proposed as being at the origin of various aspects of solar activity, they have been widely dismissed as being too feeble to cause any known physical effect. A detailed discussion, and the possibility of tests using the barycentric motion of exoplanet host stars, is given by \citet{2011A&A...525A..65P}.

There is, meanwhile, a growing appreciation that dark matter in the Galaxy may be in the form of streams \citep[e.g.][]{2011MNRAS.413.1419V,2012PDU.....1...50K}, and that these streams may provide a possible route to the directional detection of dark matter, such as WIMPS. For example, \citet{2014PhRvD..90l3511O} analysed the signal expected from a Sagittarius-like stream according to stream speed and direction.

The possible influence of dark matter on exoplanets has recently been considered by \cite{2021PhRvL.126p1101L}.  They suggest that dark matter with masses above $\sim1$~MeV can be probed through dark matter-induced exoplanet heating inversely correlated with the Galactic radial position. 
In addition, an argument for the correlation of some features of solar activity with the orbiting planets, due to the planetary lensing of streaming dark matter, has been suggested by \citet{2017PDU....17...13B}. They suggested that the existence of putative streams of dark matter, with speeds around $10^{-4}-10^{-3} c$, could then explain the puzzling behaviour of the active Sun. 
The recently observed planetary relationship of annual stratospheric temperature excursions \citep{2020PDU....2800497Z} motivates further investigations along these lines.

We propose here two tests in exoplanet systems to investigate this suggestion. 

(A) Many of the known exoplanetary systems comprise planets in the mass range of Jupiter, and over a wide range of semi-major axis (periods of a few days up to tens of years), and therefore covering a wide range of gravitational focal distance (and thus also dark matter velocity). If at least part of the solar activity is phased with the direction of one or more dark matter streams, and assuming that such a stream has a cross-section of at least a few tens of parsec, then nearby exoplanet systems might be expected to show similar host star activity possibly at the same phase in Galactocentric coordinates.  At the same time, we are not aware of exoplanet systems where there is obvious correlation between stellar activity and planetary orbital phase, other than for very close-in orbits. 

The approach would require the selection of appropriate exoplanet systems, based on their orbital configuration, and the orientation of their orbital plane with respect to the hypothesised dark matter stream(s), then examining any correlation between the two. On the assumption that the same or some other dark matter stream might affect nearby exoplanet systems in a similar manner as it does for the solar system, a Bayesian approach might be well suited.

(B) If at least part of the solar variability is indeed detecting the signature of incident streaming dark matter, and if the dark matter stream correlates with the path of ancient stellar streams, then there could be evidence in the Gaia catalogue for correlated space motion of low-metallicity (i.e. ancient) stars following this same dark matter stream. 
Broadly, this test might proceed by
(i)~selecting a subset (out of the very large number of Gaia stars) with a very low-metallicity (viz.\ very old stars, i.e. those expected to be present in these dark matter streams);
(ii) for these stars, recognising the signature of a dynamical stream by deriving the angular momentum of these stars, and searching for subsets with a common angular momentum, i.e.\ objects corresponding to a stream of stars with a certain Galactocentric impact parameter.


\end{document}